\documentclass[preprint]{aastex631}

\shorttitle{Staircase $\delta$ Scuti} \shortauthors{Felvus \& Hintz}

\begin{document}

\title{Discovery of Staircase $\delta$ Scuti Variables}

\correspondingauthor{Eric Hintz}
\email{hintz@byu.edu}

\author[0009-0003-0113-8407]{Sara Felvus}
\affiliation{Department of Natural, Health, and Mathematical Sciences, MidAmerica Nazarene University, Olathe, Kansas 66062}

\author[0000-0002-9867-7938]{Eric G. Hintz}
\affiliation{Department of Physics and Astronomy, Brigham Young University, N283 ESC, Provo, UT 84602}

\begin{abstract}
Analysis of the previously classified delta Scuti variable star MW Camelopardalis using data from the Transiting Exoplanet Survey Satellite (TESS) sparked a deeper inquiry due to unexpected patterns within the target’s observed-calculated (O-C) graph. From the shape of the O-C diagram we have designated these objects as Staircase $\delta$ Scuti. The pattern was found to be replicated in the O-C graphs of seven additional targets. The objects are TIC 17931346, TIC 44845403, TIC 123580083, TIC 173503902, TIC 302394816, TIC 194944219, and TIC 396465600.  The $Q$ value for the targets, their position in the $\delta$ Scuti Leavitt Law, and location in the instability strip would show these objects to be low mass, fundamental pulsators, near the red edge of the instability strip. We also discuss the impact this phenomenon could have on the analysis of all pulsating variable stars.
\end{abstract}


\section{Introduction}
As part of a program to provide long-term monitoring of $\delta$ Scuti variables to look for evolutionary effects, we examined the star MW Camelopardalis (MW Cam). MW Cam was discovered to be variable by the Hipparcos satellite as reported by \citet{koen02}. The star was then classified as a $\delta$ Scuti by \citet{vidal02} with two clear frequencies of 7.5301 c d$^{-1}$ and 7.8003 c d$^{-1}$. Since that point the star has received very little attention. To supplement data from our local telescopes we used the LightKurve software to obtain three epochs of data from the Transiting Exoplanet Survey Satellite (TESS). From this data we were also interested to see any short term variations in the O-C diagram (Observed - Calculated).

From an O-C diagram one would normally expect to see a straight line for a constant period, a parabola for a constant period change, or perhaps a sine curve showing an orbit. When the first O-C diagram for MW Cam was examined it surprisingly showed what is best described as a saw-tooth pattern. Adjusting the period, the pattern became a staircase. From a significant literature search no examples of this type of behavior could be found. A search of the TESS archives yielded 7 additional example of stars with a step structure in their O-C diagram. We call these objects Staircase $\delta$ Scuti. In this paper we will examine the discovery process for MW Cam, detail the other seven Staircase examples, and finally provide a summary of important characteristics for this new sub-class of $\delta$ Scuti variables.

\section{Discovery of First Staircase $\delta$ Scuti}
\begin{figure}
\figurenum{1} \epsscale{0.75} \plotone{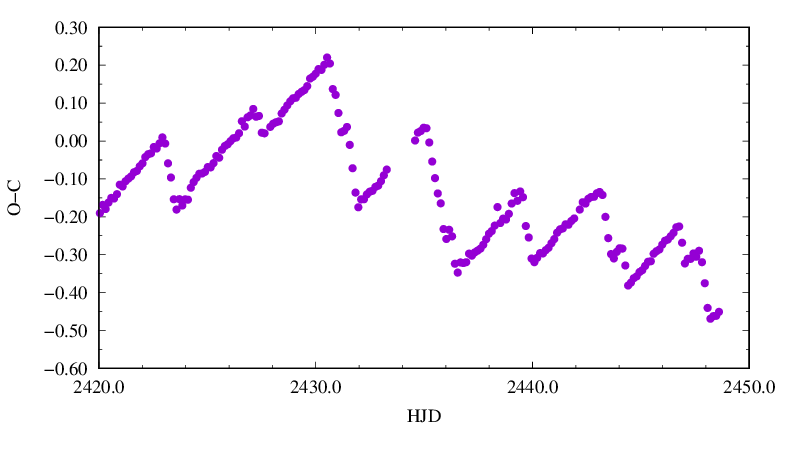}
\caption{MW Cam O-C with the unexpected `saw-tooth' pattern.\label{fig01}}
\end{figure}

From three different epochs of TESS data we obtained 619 times of maximum light for MW Cam, with 215, 208, and 196 from each epoch respectively. Times of maximum light were determined using the extrema calculator in the Peranso software package \citep{pv16}. With the almost continuous data provide by TESS, we were able to easily assign cycle numbers to each maximum. A linear fit was found and calculated times were determined for each maximum. When the O-C diagram was examined the pattern was very surprising, it displayed a saw-toothed pattern that repeated as shown in Figure~\ref{fig01}. Since we were seeing two linear relations we realized that this represented two distinct periods. The pattern was seen in all three epochs of data. However, given the pattern there was no way to determine the correct cycle numbers to bridge the time gaps.

\begin{figure}
\figurenum{2} \epsscale{0.75} \plotone{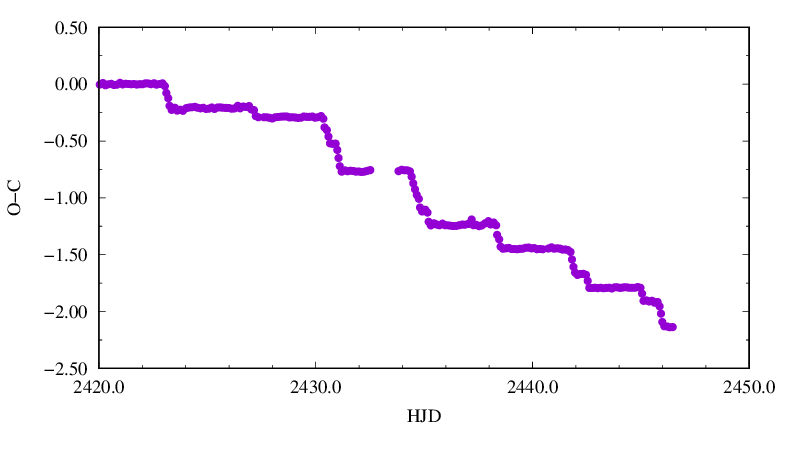}
\caption{Adjusted O-C graph for MW Cam using period from first upward sloped section. \label{fig02}}
\end{figure}

We used the times of maximum light from the first upward slanted line segment in Figure~\ref{fig01} to determine a new starting period. Using this period (matching a frequency of 7.622 c d$^{-1}$) we recalculated the O-C values and obtained the curve seen in Figure~\ref{fig02}. This generated a pattern with a set of flat parallel lines which would have the same period. Between these flat sections we found a set of roughly parallel diagonal lines connecting the flats. Therefore, two distinct periods.   

\begin{figure}
\figurenum{3} \epsscale{0.5} \plotone{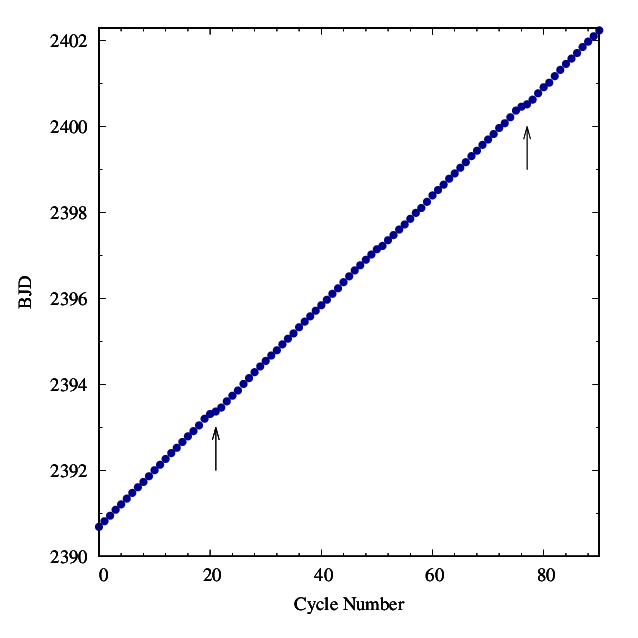}
\caption{MW Cam cycles vs observed times of max light for initial ephemeris calculation. Arrows mark times of O-C steps. \label{fig03}}
\end{figure}

Returning to the original plot of BJD versus cycle number we noted very subtle shifts as shown in Figure~\ref{fig03}. The arrows mark the points where there is a slight deflection in the linear track. Taken over the entire run of times of maximum light this makes the overall line shallower, thus underestimating the period. 

\begin{figure}
\figurenum{4} \epsscale{0.75} \plotone{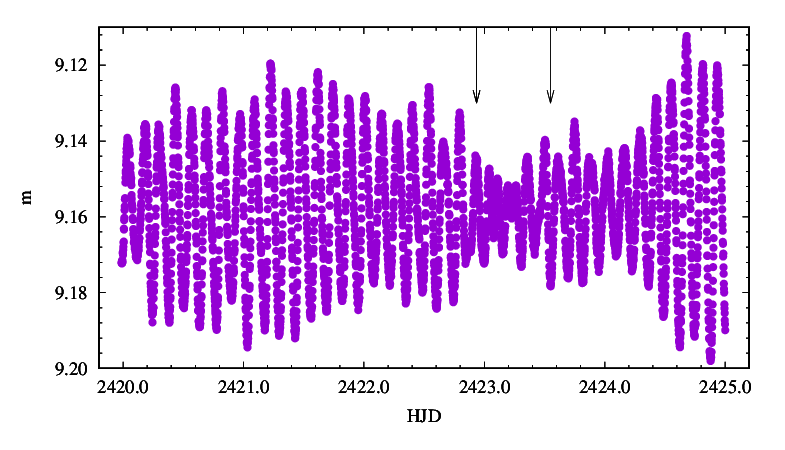}
\caption{Portion of second epoch of MW Cam TESS data. Arrows mark the sides of the transition time. \label{fig04}}
\end{figure}

In Figure~\ref{fig04} we show the early portion of the second epoch of TESS data (BJD 24572319.99 to 24572446.58). This includes data for the first flat (before first arrow), the short diagonal (between the arrows), and then a portion of the second flat. This pattern clearly shows a beat frequency that leads to a period of almost complete destructive interference when another period determines the times of maximum light. Clearly the pattern is made up of a number of pulsational periods.

\begin{figure}
\figurenum{5} \epsscale{0.75} \plotone{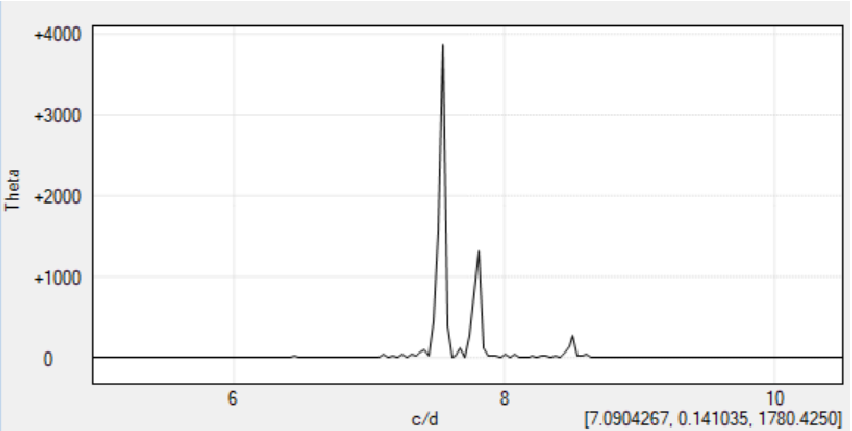}
\caption{Three frequencies found in one epoch of MW Cam data. \label{fig05}}
\end{figure}

In addition to the O-C analysis, we used the period determination packages Period04 \citep{lenz05} and Peranso (CLEANest). After a careful examination we found that both packages yielded nearly identical results. The first period search for MW Cam is shown in Figure~\ref{fig05}. This shows two close frequencies at 7.5306 c d$^-1$ and 7.7976 c d$^-1$, with two additional frequencies at 8.497 c d$^-1$ and 12.7176 c d$^{-1}$. We note that the first two frequencies match very well with those in \citet{vidal02}. In the folded light curve from \citet{vidal02} for MW Cam you can see the wide scatter in the light curve indicative of the presence of multiple frequencies.

\section{Additional Staircase $\delta$ Scuti}
Given the unique nature of the variations in MW Cam we searched for additional examples of Staircase $\delta$ Scuti. Using the Simbad Astronomical Database we searched for stars previously classified as $\delta$ Scuti, with apparent magnitudes brighter than 13th in $V$, and with fewer than 3 prior publications. This initial search resulted in a list of 780 potential targets. Given that the pattern would likely only be seen in long runs of TESS data, we pruned the list by including only targets observed by TESS. 

Two primary characteristics were present in the MW Cam data. First there was the strong beat frequency as shown in Figure~\ref{fig04} with a trough in the middle where the curve becomes more erratic and loses shape. The second feature was the pattern seen in the period search from Peranso (see Figure~\ref{fig05}). The pattern expected from the analysis of MW Cam is two closely spaced frequencies with perhaps a third offset frequency. The ratio of the two primary frequencies is near one. This search yielded 10 additional Staircase candidates that required more detailed examinations. From those 10 targets, 7 were eventually found to be Staircase $\delta$ Scuti. In Table~\ref{tbl01} we gather information about the 8 final targets including three with variable star names, the TESS cadence and number of times of maximum light for each target.

\begin{deluxetable}{rccccrcr}
  \tablecolumns{7}
  \tablecaption{Frequencies and Other Information for Staircase Targets. \label{tbl01}}
  \tablewidth{0pt}
  \tablehead{
	\colhead{TIC \#} & \colhead{Names} & \colhead{Cadence} & \colhead{ \# Max} &\colhead{Fourier} & \colhead{Frequency} & \colhead{Other} & \colhead{Frequency}\\
	\colhead{} & \colhead{} & \colhead{(min)} & \colhead{} & \colhead{} & \colhead{c d$^{-1}$} & \colhead{} & \colhead{c d$^{-1}$}}
  \startdata
        160298091 & SC 1   & 2.0  & 619 & $f_1$ & 7.5306  & Flat  & 7.621 \\
                  & MW Cam &      &     & $f_2$ & 7.7970  & Drop  & 12.730 \\
                  &        &      &     & $f_3$ & 8.4967  & Total & 8.059 \\
                  &        &      &     & $f_4$ & 12.7117 &       &       \\
                                                                     \\
        17931346  & SC 2   & 30.0 & 159 & $f_1$ & 14.0291 & Flat  & 14.105 \\
                  & MP UMa &      &     & $f_2$ & 14.2574 & Rise  & 13.837 \\
                  &        &      &     & $f_3$ & 18.0588 & Total & 14.033 \\
                  &        &      &     & $f_4$ & 20.0696 &       &        \\
                                                                     \\
        44845403  & SC 3   & 30.0 & 141 & $f_1$ & 6.5631  & Flat  & 6.458 \\
                  &        &      &     & $f_2$ & 6.2298  & Drop  & 6.655 \\
                  &        &      &     & $f_3$ & 6.4826  & Total & 6.565 \\
                  &        &      &     & $f_4$ & 10.5175 &       &       \\
                                                                    \\
        123580083 & SC 4     & 30.0 & 145 & $f_1$ & 7.1727  & Flat  & 7.144 \\
                  &          &      &     & $f_2$ & 7.0771  & Drop  & 8.230 \\
                  &          &      &     & $f_3$ & 14.2496 & Total & 7.239 \\
                                                                            \\
        173503902 & SC 5     & 30.0 & 192 & $f_1$ & 9.1633  & Flat  & 9.234 \\
                  &          &      &     & $f_2$ & 9.3689  & Rise  & 8.857 \\
                  &          &      &     & $f_3$ & 9.1585  & Total & 9.164 \\
                  &          &      &     & $f_4$ & 15.4051 &       &      \\
                                                                           \\
        302394816 & SC 7     & 30.0 & 106 & $f_1$ & 6.7096  & Flat  & 6.780 \\
                  &          &      &     & $f_2$ & 7.0176  & Rise  & 6.618 \\
                  &          &      &     & $f_3$ & 6.6290  & Total & 6.709 \\
                  &          &      &     & $f_4$ & 13.8451 &       &       \\
                                                                            \\
        194944219 & SC 9     & 10.0 & 201 & $f_1$ & 7.9068  & Flat  & 7.779 \\
                  &          &      &     & $f_2$ & 7.6973  & Drop  & 8.281 \\
                  &          &      &     & $f_3$ & 8.1282  & Total & 7.915 \\
                  &          &      &     & $f_4$ & 12.3097 &       &       \\
                                                                            \\
        396465600 & SC 11    & 10.0 & 196 & $f_1$ & 7.5667  & Flat  & 7.603 \\
                  & V366 Aps &      &     & $f_2$ & 7.6724  & Drop  & 7.726 \\
                  &          &      &     & $f_3$ & 15.2380 & Total & 7.747 \\
                  &          &      &     & $f_4$ & 7.7827  &       &       \\
    \enddata
\end{deluxetable}

\begin{figure}
\figurenum{6} \epsscale{1.0} \plotone{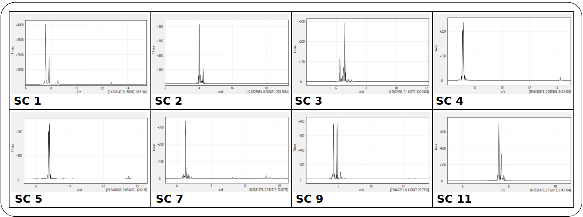}
\caption{Results of Fourier analysis for all 8 Staircase targets. \label{fig06}}
\end{figure}

A first examination of the 8 targets used Fourier analyzed similar to the one mentioned above for MW Cam. The periodogram from Peranso is shown for each of the targets in Figure~\ref{fig06}. Using Period04, we were able to determine the signal to noise ratio (SNR) for each frequency and only kept those with SNR values greater than 6. These values are gathered in column six of Table~\ref{tbl01}. Beyond the two closely spaced values for $f_1$ and $f_2$ no clear pattern was seen for all 8 objects. We do note that for SC 7 and SC 11 we found a frequency that was approximately $f_1 + f_2$.   

\begin{figure}
\figurenum{7} \epsscale{1.0} \plotone{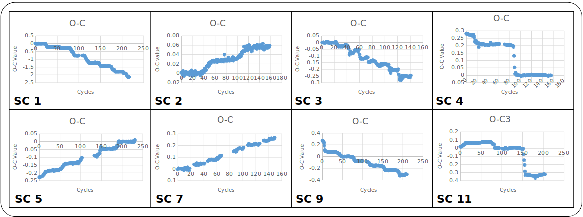}
\caption{The O-C graphs of all 8 Staircase targets. \label{fig07}}    
\end{figure}

The 7 newly found targets were processed through Peranso to locate times of maximum light and observe the O-C just as the original target. Starting with a frequency, or period, near the value of $f_1$ we generated an O-C diagram for each target. As seen in Figure~\ref{fig01} for MW Cam there are flat regions with a large number of points and then a different flat region that is made of fewer points. Using the higher density line segments, we determined a linear fit where the period was the slope. Averaging over a number of these sections we got the values labeled as Flat in column 8 of Table~\ref{tbl01}. This period was used to generate the O-C diagrams shown in Figure~\ref{fig07}. Following a similar procedure for the lower density lines we got the periods for either the Drop or the Rise of each object. The Drop/Rise lines are not nearly as well determined as the Flat lines and therefore likely have a larger error. We determined a final period from a linear fit to the entire run of times of maximum light. These are included in Table~\ref{tbl01} as a frequency labeled as Total. 

One of the most difficult questions to answer is which frequency is the actual pulsation of the star. Given that the Flat frequency provides the best phasing of the data we feel that this is the true pulsational frequency. We also note that in each case the frequency from the Flat is between $f_1$ and $f_2$. For MW Cam we have $f_4$ being very close in value to the Drop value. However, this is only true for MW Cam. We are concerned that some values could be impacted by the longer cadence in all targets except MW Cam. One final point is that if Flat is larger than Drop/Rise we have an upward staircase and if Drop/Rise is larger than Flat we have a downward staircase.

For completeness we wish to address the remaining 3 candidates from the initial search. We eliminated TIC 23652761 from further consideration because the cadence of observations were not sufficient to provide good measurements for the times of maximum light. This target should be investigated with a higher cadence rate. TIC 21509603 didn't show the step pattern seen in the other targets. The published period for this star is 0.057525 d, which might also require a higher cadence. Finally, the O-C diagram from TIC 375082030 was jumbled and didn't show a clear Staircase pattern. 

\section{Characterization of Staircase $\delta$ Scuti}
\subsection{Fundamental pulsations and a Leavitt Law relation}
To put these stars in context we gathered information from Gaia DR3 and the TESS Input Catalog on the basic characteristics of the stars. This included values for $T_{eff}$, $\log (g)$, and distance. Using this information and extinction values from the GALExtin maps \citep{amores21}, we determined the absolute magnitude for each object. Finally, we used the information in \citet{breger75} to determine the pulsational constant $Q$. This information is all gathered in Table~\ref{tbl02}. The $Q$ values would seem to indicate that all these targets are pulsating in the fundamental frequency. This is also supported by plotting the absolute magnitude versus period. In Figure~\ref{fig08} we show all 8 stars on a Leavitt Law relation (PL). All targets follow the fundamental PL relation from \citet{poro21}, with none near the first overtone line. We converted the absolute magnitude to a value in solar luminosity and combined this with the $T_{eff}$ value to examine the location of the stars in the \citet{netzel22} instability strips. We found that all targets are to the red edge of the instability strip and in regions considered to be populated by fundamental pulsators. 

\subsection{Spectral data}
\begin{figure}
\figurenum{8} \epsscale{0.6} \plotone{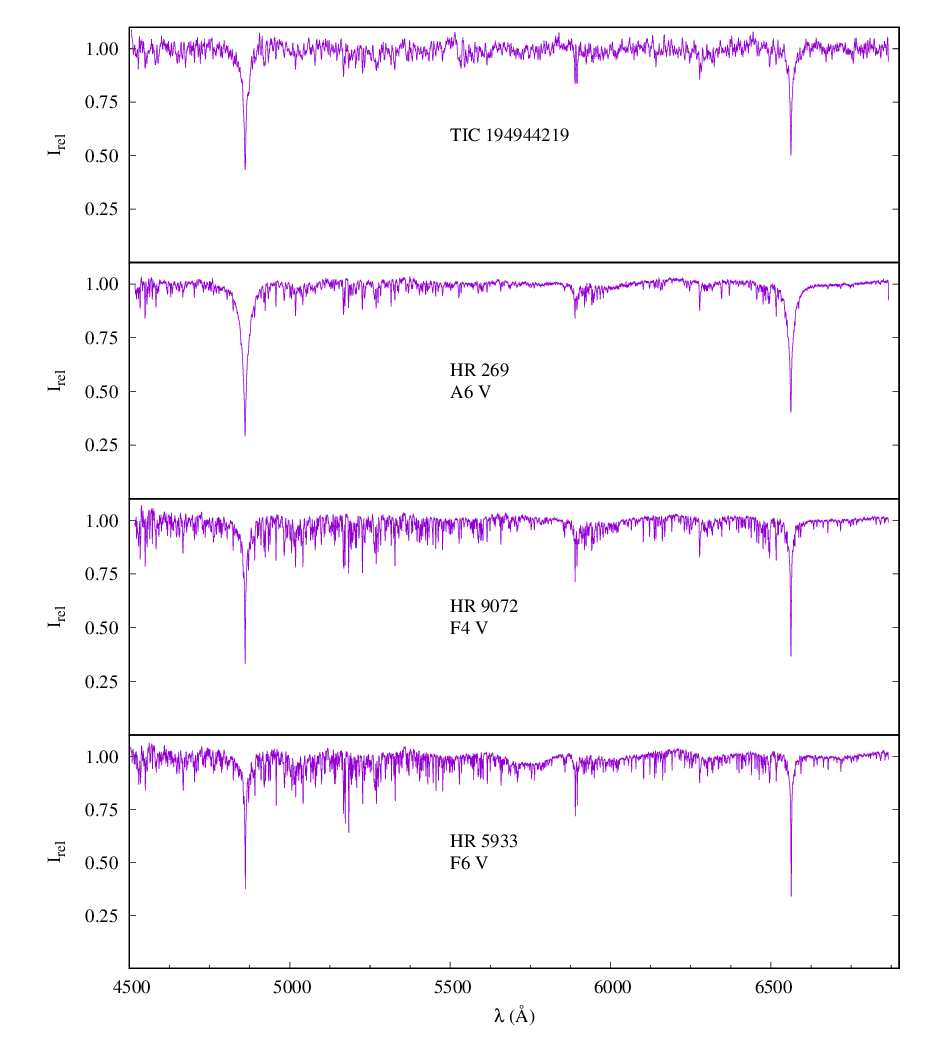}
\caption{The O-C graphs of all 8 Staircase targets. \label{fig08}}    
\end{figure}

Using the 1.2-m Telescope at the Dominion Astrophysical Observatory we obtained 8 spectra of TIC 194944219 over 4 nights in July 2024. The spectrograph was configured similar to that reported in \citet{joner15}. We used the calibrations in \citet{joner15} to determine an H-alpha index of $2.759\pm0.005$ which would indicate roughly a F3 spectral type. In Figure~\ref{fig08} we show the spectrum of TIC 194944219 compared to 3 standard stars from \citet{joner15} and again estimate an F3 V classification. The location in the instability strip discussed above fits well with this designation.

\subsection{Comparison of Parameters}
An examination of Table~\ref{tbl02} shows a number of characteristics for the final set of 8 Staircase $\delta$ Scuti variables. First we note that they are spread across a full range of galactic longitudes, however 6 of the 8 targets are near the galactic plane. The average mass for the group is $1.6\pm0.1$ M$_\sun$. Given that the expected mass range for solar mass $\delta$ Scuti is 1.5 to 2.5 M$_\sun$, these are mostly low mass members of the group.

\begin{deluxetable}{lrrccrccc}
\tabletypesize{\scriptsize}
  \tablecolumns{8}
  \tablecaption{Information from Gaia EDR3 and TESS Input Catalog. Plus Determined $M_{V}$ and Q values.\label{tbl02}}
  \tablewidth{0pt}
  \tablehead{
	\colhead{Target} & \colhead{$l$} & \colhead{$b$} & \colhead{$T_{eff}$} & \colhead{$\log{g}$} & \colhead{Mass} & \colhead{distance} & \colhead{$M_{V}$} & \colhead{Q}\\
	\colhead{} & \colhead{$\arcdeg$} & \colhead{$\arcdeg$} & \colhead{K} & \colhead{} & \colhead{M$_\sun$} & \colhead{pc} & \colhead{}& \colhead{}}
  \startdata
    TIC 160298091 & 124.05617 & 35.60029 & 7240 & 3.65 & 1.63 & 341 & 1.26 & 0.033\\
    TIC 17931346  & 174.24160 & 67.52000 & 7800 & 4.12 & 1.85 & 775 & 2.11 & 0.039\\
    TIC 44845403  & 268.67029 & 10.30611 & 6780 & 3.41 & 1.45 & 468 & 1.11 & 0.026\\
    TIC 123580083 & 220.31264 & -3.58700 & 6870 & 3.55 & 1.49 & 1216 & 1.07 & 0.028\\
    TIC 173503902 & 335.22337 & -8.51828 & 6930 & 3.71 & 1.51 & 458 & 1.71 & 0.030\\
    TIC 302394816 & 140.95510 & -8.02084 & 7300 & 3.62 & 1.65 & 1130 & 0.99 & 0.033\\
    TIC 194944219 &  80.11326 & -3.99887 & 7220 & 3.61 & 1.62 & 719 & 1.27 & 0.030\\
    TIC 396465600 & 308.62187 &-12.30560 & 6760 & 3.55 & 1.45 & 741 & 1.59 & 0.029
  \enddata
\end{deluxetable}

\begin{figure}
\figurenum{9} \epsscale{0.6} \plotone{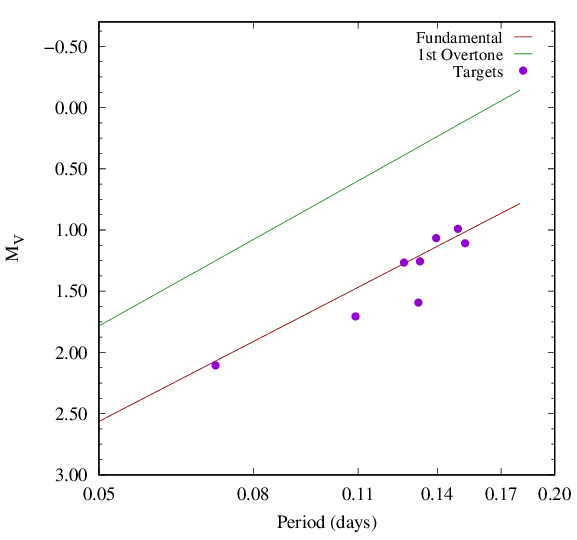}
\caption{All 8 Targets on Leavitt Law relation with the Fundamental and 1st Overtone relations from \citet{poro21} included. \label{fig09}}
\end{figure}

\section{Conclusions}
From an analysis of 8 targets from the TESS data archive we report the discovery of a new sub-class of $\delta$ Scuti variable stars: the Staircase $\delta$ Scuti. These objects are defined by a staircase pattern found in their O-C diagrams that is made of two frequencies. We note that the primary frequency from a Fourier analysis doesn't match frequencies found O-C analysis. 

An examination of the basic characteristics of the 8 objects studied show some patterns. We find that the majority of the targets are on the low mass side of the $\delta$ Scuti mass range and to the red edge of the instability strip. From the $Q$ value and the stars location on the PL relation we classify all targets as pulsating in the fundamental mode. 

The discover of the Staircase $\delta$ Scuti opens a lot of questions about using O-C observations to examine long term period changes in these type of objects. Could a staircase pattern mimic a parabolic O-C shape if the measurements of times of maximum light are too spread out? Could these patterns be seen in stars that have been monitored and show sharp period changes like those reported for DY Pegesi \citep{berthold24}? How long can the Flat portion continue before a drop or rise? There are many questions that need to be address for these new objects and their impact on our interpretation of O-C diagrams.

\section{Acknowledgements}
\begin{acknowledgments}
 Based on observations obtained at the Dominion Astrophysical Observatory, Herzberg Astronomy and Astrophysics Research Centre, National Research Council of Canada. We also would like to acknowledge National Science Foundation grant \text{\#}2348770 for support of this research.

This work has made use of data from the European Space Agency (ESA) mission
{\it Gaia} (\url{https://www.cosmos.esa.int/gaia}), processed by the {\it Gaia}
Data Processing and Analysis Consortium (DPAC,
\url{https://www.cosmos.esa.int/web/gaia/dpac/consortium}). Funding for the DPAC
has been provided by national institutions, in particular the institutions
participating in the {\it Gaia} Multilateral Agreement.

This paper includes data collected with the TESS mission, obtained from the MAST data archive at the Space Telescope Science Institute (STScI). Funding for the TESS mission is provided by the NASA Explorer Program. STScI is operated by the Association of Universities for Research in Astronomy, Inc., under NASA contract NAS 5–26555.
\end{acknowledgments}

\facilities{TESS,DAO:1.22m}

\end{document}